# Distributed Apportioning in a Power Network for providing Demand Response Services

Sourav Patel, Sandeep Attree, Saurav Talukdar, Mangal Prakash, Murti V. Salapaka
Department of Electrical and Computer Engineering
University of Minnesota
{patel292, attree002, taluk005, praka027, murtis}@umn.edu

*Abstract*—Greater penetration of Distributed Energy Resources (DERs) in power networks requires coordination strategies that allow for self-adjustment of contributions in a network of DERs, owing to variability in generation and demand. In this article, a distributed scheme is proposed that enables a DER in a network to arrive at viable power reference commands that satisfies the DERs local constraints on its generation and loads it has to service, while, the aggregated behavior of multiple DERs in the network and their respective loads meet the ancillary services demanded by the grid. The Net-load Management system for a single unit is referred to as the Local Inverter System (LIS) in this article . A distinguishing feature of the proposed consensus based solution is the distributed finite time termination of the algorithm that allows each LIS unit in the network to determine power reference commands in the presence of communication delays in a distributed manner. The proposed scheme allows prioritization of Renewable Energy Sources (RES) in the network and also enables auto-adjustment of contributions from LIS units with lower priority resources (non-RES). The methods are validated using hardware-in-the-loop simulations with Raspberry PI devices as distributed control units, implementing the proposed distributed algorithm and responsible for determining and dispatching real-time power reference commands to simulated power electronics interface emulating LIS units for demand response.

## 1. INTRODUCTION

The existing power grid network is in the process of transitioning from a framework, where conventional large power plants generate at one end while loads consume at the other, towards integration of large number of smaller distributed generation (DG) units scattered throughout the network to provide ancillary services and support the power grid [1], [2], [3]. The coordination of multiple distributed generation units presents significant challenges. [4] and [5] provide a review of works that have employed a centralized approach to solve the problem of aggregating and coordinating Distributed Energy Resources (DERs), where a secondary centralized controller dispatches commands to DERs and requires information from all of the DERs, in many cases is not tenable. A distributed method for grid ancillary services has a number of advantages over centralized architecture as enumerated below:

1) Distributed co-ordination is achieved using only local computations (which reduces communication overheads and congestion) pertaining only to the neighborhood of the DER.
2) Distributed coordination facilitates 'plug and play' capability, where a new DER that needs to connect to the network requires communication only to its nearest available DER.
3) Distributed architecture is resilient to failures as any particular node failure does not lead to failure of the entire network.

The objective of this article is to meet the grid ancillary demand (global objective) by using an aggregation of DERs forming a network while respecting their local generation and demand constraints in a distributed manner. In this regard, we propose a distributed resource apportioning framework that is suitable for real-time implementation. We envision the use of ultra low-cost computing devices like Raspberry Pi (R-Pi) acting as control and communication agents in the DER network. The wireless communication channel between R-Pis suffers from time varying (stochastic) delays, which for practical purposes can be considered to be bounded (see Figure 5(b)). The distributed resource apportioning solution proposed here is robust to the presence of bounded communication delays between the DERs' communication devices.

Ratio consensus algorithm for distributed coordination of DERs to meet the ancillary service demand is presented in [6]. The ratio consensus algorithm is used to compute the power reference commands for the DERs based on their generation capacities. However, the convergence of ratio consensus to the power reference command values is asymptotic in nature and hence, unsuitable for real time implementation. To circumvent this issue, a distributed finite time termination of ratio consensus is presented in [7], which builds on ideas presented in [6] and [8] for frequency regulation in a network of islanded AC micro grids. However, none of the works discussed above address the issue of communication delays in ratio consensus and the method for distributed termination is not present. Distributed finite time termination of ratio consensus in the presence of bounded delays is presented in [9] but it does not adhere to distributed resource apportioning and coordination of DERs in the presence of communication delays. Figure 1 shows experimental results for implementing ratio consensus algorithm on a 5 node network for averaging of initial values without accounting for any communication delays. Ignoring communication delays results in a converged value of 225, that is, an error of $43.75\%$ from the true average of 400. This motivates the need to address the challenges of communication delays for successful implementation of distributed consensus algorithms for real-time applications such as power networks. Furthermore, the power network relies on a number of measurements and communicating these measurements among control units for purposes of effective control [10]. Ignoring communication delays could result in a destabilizing effect on the entire system or any part thereof; hence, it is important that communication delays are taken care of for meeting performance specifications.

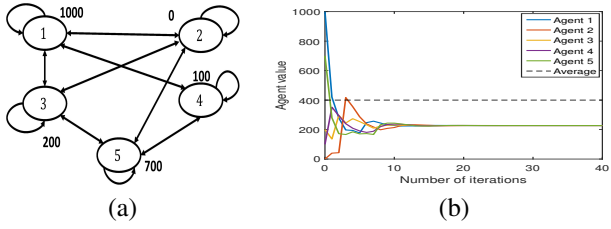

Fig. 1. (a) 5 node network with initial conditions, (b) experimental results for the 5 node network when communication delays are not considered converge at a value different than the true average.

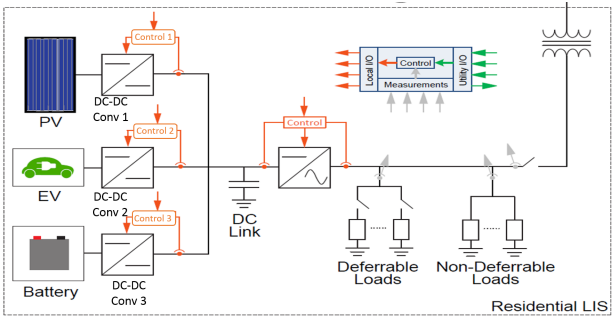

Fig. 2. Residential scale LIS unit with generation sources and local loads

In this article we present a distributed scheme for apportioning the share of resources that a DER has to provide toward meeting a global demand while satisfying local load demands. The effectiveness of the algorithm is demonstrated by executing the algorithm on real devices (Raspberry Pis) where the power reference commands, obtained from the termination of ratio consensus, are implemented in a hardware in the loop simulation with Simulink. We use a number of basic notions from graph theory and linear algebra which are essential for the subsequent developments. Detailed description of these notions are available in [11] and [12].

In the next section we provide an overview of the Local Inverter System (LIS), the associated network topology as well as the resource apportioning problem.

## 2. LOCAL INVERTER SYSTEMS

### A. System Topology

A residential level *local inverter system* (Figure 2) consists of an array of distributed generation sources (such as PV, EV, battery) interfaced with DC-DC converters, DC Link Capacitor and DC-AC inverters to serve local AC loads. Utility scale LIS unit consists of an utility scale inverter with a PV array to meet ancillary demand services of the grid and/or a network of residential LIS units. The output of an LIS unit is connected through a point of common coupling (PCC) to other LIS units in a peer-to-peer microgrid or to the grid through an aggregator (described later) when in grid-connected mode.

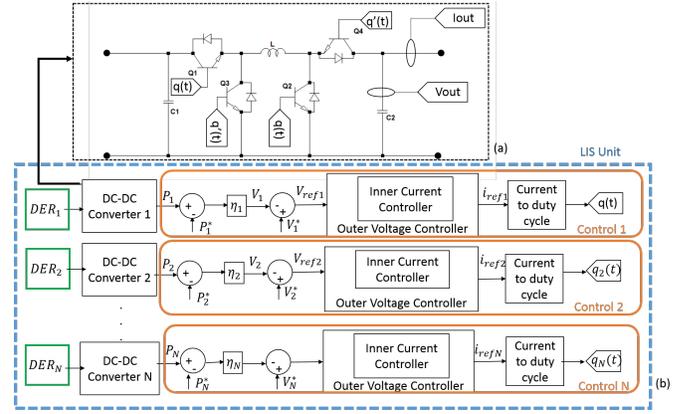

Fig. 3. (a) Bidirectional DC-DC Converter realizations for DERs in a Residential LIS unit where q(t) is the desired duty ratio input, (b) A single LIS unit with $N$ DERs implementing power-voltage droop control and inner-control loops for voltage regulation and load sharing.

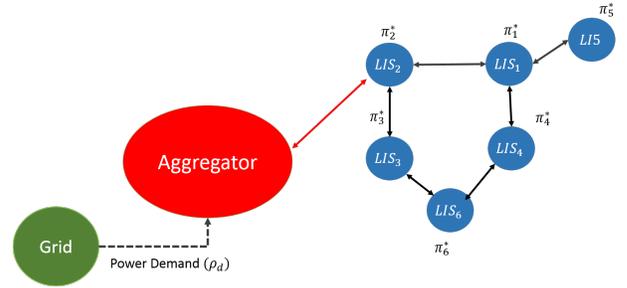

Fig. 4. Aggregator injecting grid ancillary demand $\rho_d$ at node LIS 2 in a 6-node LIS network.

Consider a single LIS unit with $N$ DERs connected in parallel. A power-voltage droop control is implemented for each DC-DC converter unit as described in [13] and extended to the parallel converter system of Figure 3. The Power output of $j^{th}$ DC-DC converter, $P_j$, is compared against a desired reference power output set point, $P_j^*$, demanded from the $j^{th}$ DER. A suitable droop coefficient $\eta_j$ is chosen that generates voltage reference set point $V_{ref_j}$ according to equation (1) for the inner loop controller. Thus,

$$V_{ref_j} = V^* - \eta_j(P_j - P_j^*). \qquad (1)$$

The inner-loop controller consists of a voltage controller loop that generates a current reference signal $i_{ref_j}$ and a fast current controller loop for voltage regulation of the $j^{th}$ DC-DC Converter. An extensive design of such inner-loop controller is proposed in [14] that implements a decentralized control architecture for load sharing among multiple parallel converters similar to the topology proposed here. The converter devices are assumed to be lossless in this article. The reactive power flow is assumed to be zero in the network and transmission lines are considered to be lossless.

### B. Network of LIS units

An *Aggregator* as shown in Figure 4 is an entity that injects the grid ancillary demand into the network of LIS units, which then becomes the global objective of the network. Consider a network of $\mathcal{I} = \{1, 2, ..., n\}$ LIS units, where, the $j^{th}$ LIS unit has $E_j$ DERs and $L_j$ loads. The generation capacity of the $j^{th}$ LIS unit, $\pi_g^j := \sum_{m=1}^{E_j} \pi_{g_m}^j$, where, $\pi_{g_m}^j$ is the generation capacity of the $m^{th}$ DER in the $j^{th}$ LIS unit. The local load demand of the $j^{th}$ LIS unit, $\pi_l^j := \sum_{m=1}^{L_j} \pi_{l_m}^j$, where, $\pi_{l_m}^j$ is the demand of the $m^{th}$ load in the $j^{th}$ LIS unit. At a given instant, the $j^{th}$ LIS unit is said to have a net reserve $R_j$, where,

$$R_j = \sum_{m=1}^{E_j} \pi_{g_m}^j - \sum_{m=1}^{L_j} \pi_{l_m}^j \text{ for all } j = 1, 2, ..., n. \quad (2)$$

If the net reserve $R_j > 0$, LIS unit $j$ behaves as a network source otherwise it is a network load. In this article, we assume all LIS units act as sources to meet the grid demand $\rho_d$; extending the resource apportioning framework to the case where there is a mix of source and load LIS units is a simple extension.

The LIS units communicate with each other as dictated by the network topology through wireless channels. This communication link is represented as a bidirectional edge in the network and results in an undirected graph. Delays are inherent because of the presence of wireless communication channels between control units of LIS units, and are assumed to be fixed and uniformly bounded for simplicity of presentation and the HIL validation is for the stochastic but bounded delay case.

### C. Resource Apportioning Problem

The network of LIS units can be viewed as a multi agent system with the agents interacting with their neighbors over a communication network. Each LIS unit is considered as a node and each communication link between LIS units is considered as an undirected edge in a graph $G = (V, E)$, where, $V$ denotes the set of nodes and $E$ denotes the set of edges. The central aggregator requests that a total resource $\rho_d$ be supplied, which has to be collectively provided by the $n$ LIS units, while respecting their individual resource constraints. Let $\pi_i^*$ be the amount of resource supplied by the $i^{th}$ LIS unit, where $\pi_i^{min}$ and $\pi_i^{max}$ represent the minimum and the the maximum amount of power $i^{th}$ agent can supply respectively. The objective is to determine $\{\pi_i^*\}_{i \in V}$ such that,

$$\pi_i^{min} \leq \pi_i^* \leq \pi_i^{max} \quad \text{and} \quad \sum_{i \in V} \pi_i^* = \rho_d.$$

We refer to the problem of determining $\pi_i^*$ respecting the above constraints as the resource apportioning problem. We assume that $\sum_{i=1}^n \pi_i^{min} \leq \rho_d \leq \sum_{i=1}^n \pi_i^{max}$ holds to ensure feasibility of the resource allocation problem. Let $\tau_{ij}$ denote the delay in the communication link from node $j$ to node $i$. The communication delay in each link is assumed to be constant and bounded by $\bar{\tau}$. We present below an approach which meets the demand $\rho_d$ asymptotically while respecting the constraints of each agent via a distributed iterative algorithm where each node updates its state based on its current state and the state of its neighbors.

### 3. RESOURCE APPORTIONING USING DISTRIBUTED AVERAGING

The resource apportioning problem in the presence of bounded communication delays between LIS units can be solved by using average consensus protocols which incorporates delays. We first summarize below the distributed averaging protocol in presence of bounded delays as presented in [15].

### A. Distributed Averaging Protocol

A1. Let $p_{ij}$ denote the weight on the information coming from node $j$ to node $i$. Weight matrix $[P](i, j) = p_{ij}$ associated with the undirected graph is primitive and column stochastic.

A2. The undirected graph $G = (V, E)$ is connected.

A3. Any node $i \in V$ in the undirected graph $G = (V, E)$ has access to its own value at any instant $k$ without any delay.

A4. The delay on the edge from node $j$ to $i$, $\tau_{ij}$ is bounded by some constant $\bar{\tau}$, for all $i, j \in V$, that is, $\tau_{ij} \leq \bar{\tau} < \infty$.

Under the assumptions *A1-A4* consider the iterations,

$$x_i(k+1) = p_{ii} x_i(k) + \sum_{j \in N_i^-} p_{ij} x_j(k - \tau_{ij}), \quad (3)$$

$$w_i(k+1) = p_{ii} w_i(k) + \sum_{j \in N_i^-} p_{ij} w_j(k - \tau_{ij}), \quad (4)$$

with the initial conditions be given by $x(0) = [x_1(0) \; x_2(0)...x_n(0)]^T$ and $w(0) = 1_n$ where $1_n$ is a $n \times 1$ column vector of all ones. Then the ratio of $x_i(k)$ and $w_i(k)$ asymptotically converges to $\lim_{k \to \infty} \mu_j(k) = \frac{\sum_{i=1}^n x_i(0)}{n}$ for all $j = 1, ..., n$, where $\mu_j(k) := x_j(k)/w_j(k)$ [15].

**Remark 1.** *The weight matrix being column stochastic enables the weights $p_{ij}$ to be chosen in a distributed manner. A simple scheme for choosing the weights are $p_{ji} = \frac{1}{D_i^+ + 1}$ for all $j \in \{\mathcal{N}_i^+ \cup i\}$.*

Now we use the distributed averaging based on (3) and (4) for distributed resource allocation. Suppose that the resource demanding authority, which is referred to as the aggregator can communicate to $p$ agents out of $n$ ($1 \leq p \leq n$) and send across its demand $\rho_d$. The nodes to which the aggregator relays its demand are called demand circulation nodes and the set of such nodes is denoted by $\mathcal{N}_d$. Suppose that each node has three states, $[r_i(k), s_i(k), t_i(k)]^T$ such that,

$$r_i(k+1) = p_{ii} r_i(k) + \sum_{j \in N_i^-} p_{ij} r_j(k - \tau_{ij}), \quad (5)$$

$$r_i(0) = \frac{\rho_d}{p} - \pi_i^{min} \text{ if } i \in \mathcal{N}_d \text{ or}$$

$$r_i(0) = -\pi_i^{min} \text{ if } i \notin \mathcal{N}_d,$$

$$s_i(k+1) = p_{ii} s_i(k) + \sum_{j \in N_i^-} p_{ij} s_j(k - \tau_{ij}), \quad (6)$$

$$s_i(0) = \pi_i^{max} - \pi_i^{min}, \text{ for all } i \in V,$$

$$t_i(k+1) = p_{ii} t_i(k) + \sum_{j \in N_i^-} p_{ij} t_j(k - \tau_{ij}), \quad (7)$$

$$t_i(0) = 1, \text{ for all } i \in V,$$

where, $\pi_i^{max}$ and $\pi_i^{min}$ denote the maximum and minimum power availability from the $i^{th}$ LIS unit. It follows from the above discussion that, $\lim_{k \to \infty} \frac{r_i(k)}{t_i(k)} = \frac{\rho_d - \sum_{j=1}^{n} \pi_j^{min}}{n}$, and

$$\lim_{k \to \infty} \frac{s_i(k)}{t_i(k)} = \frac{\sum_{j=1}^{n}(\pi_j^{max} - \pi_j^{min})}{n}, \text{ for all } i \in V.$$

**Theorem 3.1.** *Let the power reference command for the $i^{th}$ LIS unit be defined as,* $\pi_i^* := \pi_i^{min} + \lim_{k \to \infty} \frac{r_i(k)}{s_i(k)}(\pi_i^{max} - \pi_i^{min})$. *Then* $\sum_{i=1}^{n} \pi_i^* = \rho_d$ *and* $\pi_i^{min} \leq \pi_i^* \leq \pi_i^{max}$ *for all* $i \in V$.

*Proof.* It follows from the above discussion that,

$$\pi_i^* = \pi_i^{min} + \frac{\rho_d - \sum_{j=1}^{n} \pi_j^{min}}{\sum_{j=1}^{n}(\pi_j^{max} - \pi_j^{min})}(\pi_i^{max} - \pi_i^{min}), \text{ for all } i \in V.$$

It is clear from the feasibility of demand $\rho_d$ that, $\pi_i^{min} \leq \pi_i^* \leq \pi_i^{max}$. Moreover, it follows that,

$$\sum_{i=1}^{n} \pi_i^* = \rho_d.$$

This completes the proof. □

Theorem 3.1 provides a distributed protocol to allocate resources to meet the demand $\rho_d$. It is an extension of the ratio consensus approach for resource apportioning presented in [6] to incorporate communication delays between the agents. It should be noted that the result in Theorem 3.1 is asymptotic. Thus, the agents in principle keep updating their states (using (5),(6),(7)) forever without termination, which makes it untenable from a real time implementation perspective. One needs to terminate the computations at each node in a distributed manner when the sum of the contribution from each node is 'close' to $\rho_d$. This problem of distributed finite time termination of resource apportioning is dealt in the next section.

## 4. DISTRIBUTED FINITE TIME TERMINATION OF RESOURCE APPORTIONING

In this section, first results based on the update rules (3) and (4) are established followed by the definitions and convergence of Max-Min consensus algorithms. Subsequently, a finite-time termination criterion for average consensus is developed based on these results. Let us consider the maximum and minimum value of the ratio of consensus protocols given by (1) and (4) over all nodes within a horizon $\bar{\tau}$ from any time instant $k$ be given by,

$$M(k) := \max_{\substack{j \in V \\ r = \{0,1,2,...,\bar{\tau}\}}} \frac{x_j(k-r)}{w_j(k-r)}, w_j(k-r) \neq 0, j \in V \quad (8)$$

$$m(k) := \min_{\substack{j \in V \\ r = \{0,1,2,...,\bar{\tau}\}}} \frac{x_j(k-r)}{w_j(k-r)}, w_j(k-r) \neq 0, j \in V \quad (9)$$

**Remark 2.** *It is shown in [9] that $\{M(k)\}_{k \in \mathbb{N}}$ and $\{m(k)\}_{k \in \mathbb{N}}$ converges to $\frac{\sum_{j=1}^{n} x_j(0)}{n}$.*

**Remark 3.** *Given that $w_j(0) = 1$ for all $j \in V$ and $P$ is a non-negative matrix, $w_j(k) \neq 0$ for all $k \in \mathbb{N}$.*

Next, we introduce the Maximum Consensus and Minimum Consensus protocols.

### A. Maximum and Minimum Consensus Protocols [16]

Maximum Consensus Protocol (MXP) computes the maximum of the given initial node conditions $z(0) = [z_1(0) \, z_2(0)....z_n(0)]^T$ in a distributed manner. It takes $z(0)$ as an input and generates a sequence of node values based on the following update rule for node $i$,

$$z_i(k\bar{\tau} + q) = z_i(k\bar{\tau} + q - 1), q \in \{k+1, k+2, \cdots, k+\bar{\tau}\},$$
$$z_i((k+1)(\bar{\tau}+1)) = \max_{j \in N_i^- \cup \{i\}} z_j((k+1)(\bar{\tau}+1) - (\tau_{ij}+1)) \text{ for } k \geq 0. \quad (10)$$

The Minimum Consensus Protocol (MNP) computes the minimum of the given initial node conditions $y(0) = [y_1(0) \, y_2(0)....y_n(0)]^T$ in a distributed manner. It takes $y(0)$ as an input and generates a sequence of node values $y(k)$ based on the following update rule:

$$y_i(k\bar{\tau} + q) = y_i(k\bar{\tau} + q - 1), q \in \{k+1, k+2, \cdots, k+\bar{\tau}\},$$
$$y_i((k+1)(\bar{\tau}+1)) = \min_{j \in N_i^- \cup \{i\}} y_j((k+1)(\bar{\tau}+1) - (\tau_{ij}+1)), \text{ for } k \geq 0. \quad (11)$$

**Remark 4.** *MXP and MNP converge to the maximum and minimum of the initial conditions respectively in $D(1 + \bar{\tau})$ iterations, where $D$ is the diameter of the network* [16].

### B. Distributed Finite Time Termination of Resource Apportioning

In this section, we propose an algorithm using the MXP-MNP for stopping the ratio consensus protocol in finite time based on a user-specified threshold $\rho$. This framework first appeared in [7] as an extension of the Max-Min consensus based finite time termination of averaging consensus [8]. However, [7], [8] do not consider any communication delays in the network and [8] is based on the weight matrix $P$ being doubly stochastic thereby restrictive in the sense of distributed selection of the edge weights. A MXP-MNP framework for finite time termination of ratio consensus ((3) and (4)) with uniformly bounded communication delays and column stochastic weight matrix is reported in [9]. We apply the results in [9] for distributed finite time termination of the numerator and denominator ratio consensus protocol as shown in Algorithm 1. The initial conditions for the MXP and MNP protocols are set as the initial ratio held by the nodes, that is, $z_i(0) = r_i(0)/s_i(0)$ and $y_i(0) = r_i(0)/s_i(0)$. The MXP and MNP protocols at each node $i \in V$ are re-initialized at $k = \theta(D(1+\bar{\tau}) + \bar{\tau})$, where $\theta = 1, 2, ...,$ with $z_i(k) = \frac{r_i(k)}{s_i(k)}$ and $y_i(k) = \frac{r_i(k)}{s_i(k)}$. Given a threshold $\rho > 0$, it is proven in [9] that Algorithm 1 terminates in finite number of iterations. Note that $\{t_i\}_{i \in V}$ iterations were used for analysis only and is not required to be maintained at the nodes as its effect is cancelled.

**Algorithm 1:** Distributed finite-time termination of resource apportioning in presence of communication delays (at each node $i \in V$)

**Repeat:**
  **Input:**
    $\pi_i^{\min}, \pi_i^{\max}, \rho_d, p$
    $\rho, \bar{\tau}, D$
  **Initialize:**
    $r_i(0) = \frac{\rho_d}{p} - \pi_i^{\min}$, if $i \in \mathcal{N}_d$, else $r_i(0) = -\pi_i^{min}$;
    $s_i(0) = \pi_i^{\max} - \pi_i^{\min}$;
    $k := 0$;
    $z_i := r_i(0)/s_i(0)$;
    $y_i := r_i(0)/s_i(0)$;
    $l := 1$;
    $\theta := 1$;
  **Repeat:**
    /* ratio consensus updates of node
       i given by (3) and (4)       */
    $r_i(k+1) := p_{ii}r_i(k) + \sum_{j \in N_i^-} p_{ij}r_j(k - \tau_{ij})$;
    $s_i(k+1) := p_{ii}s_i(k) + \sum_{j \in N_i^-} p_{ij}s_j(k - \tau_{ij})$;
    **if** $k + 1 = l(1 + \bar{\tau})$ **then**
      /* maximum and minimum
         consensus updates given by
         (10) and (11) for each node
         $i \in V$                      */
      $z_i := \max_{j \in N_i^- \cup \{i\}} z_j$;
      $y_i := \min_{j \in N_i^- \cup \{i\}} y_j$;
      $l := l + 1$
    **end**
    **emit:** $r_i(k+1), s_i(k+1), y_i$ and $z_i$
    **if** $k + 1 = \theta(D(1 + \bar{\tau}) + \bar{\tau})$ **then**
      **if** $z_i - y_i < \rho$ **then**
        $r_i^* = r_i(k+1)$;
        $s_i^* = s_i(k+1)$;
        **break** ; // stop $r_i$, $s_i$, $y_i$ and $z_i$
          updates
      **else**
        $z_i :=$
          $r_i(\theta(D(1+\bar{\tau})+\bar{\tau}))/s_i(\theta(D(1+\bar{\tau})+\bar{\tau}))$;
        $y_i :=$
          $r_i(\theta(D(1+\bar{\tau})+\bar{\tau}))/s_i(\theta(D(1+\bar{\tau})+\bar{\tau}))$;
        $\theta := \theta + 1$;
      **end**
    **end**
    $k = k + 1$
  $\pi_i^* := \pi_i^{min} + \frac{r_i^*}{s_i^*}(\pi_i^{max} - \pi_i^{min})$ // power
    reference command for node $i \in V$

**Remark 5.** *The MXP and MNP iterations can be used to compute $\beta_i(\theta)$ in a distributed manner at each node. Algorithm 1 presented below uses $\beta_i(\theta) < \rho$ as a stopping criteria for termination of (5) and (6). It computes the approximate limiting ratio $r_i^*/s_i^*$ (each node converges 'close' to $\lim_{k \to \infty} r_i(k)/s_i(k)$) in a distributed manner. The only global parameters needed by each node are upper bounds on both the maximum delay and diameter of the network.*

## 5. RESULTS

Hardware in the loop (HIL) simulations are performed on a network of 6 LIS units as shown in Figure 4 for demonstrating the applicability of the algorithm presented. Each LIS unit is equipped with a Raspberry-PI 3 (Model B V1.2) (see Figure 5 (a)) [17] module for control and communication with other LIS units. R-PI 3 implements 802.11n Wireless standard that allows for a larger range of communication with faster data transmission speeds [18]. LIS units communicate in their respective neighborhoods (Figure 4) wirelessly. Figure 5 (b) shows pairwise communication delays experienced by R-PI units. These delays are due to inherent uncertainties in the communication channel as well as those modeled for the experiment.

If an LIS unit has RES, then it has a time dependent generation profile associated with it because of the variability in environmental parameters that impact generation, otherwise a fixed maximum and minimum generation capacity is assumed. For the considered topology as depicted in Figure 4, $D = 3$, $\bar{\tau} = 3$ and the weight matrix P (column stochastic) is chosen distributively as described in Remark 1. Let the grid ancillary demand, $\rho_d$ be 7000 W. In this simulation, LIS 2 is chosen to be a RES, which is a PV array operating in Maximum Power Point Tracking (MPPT) mode [19] with a suitable profile selected to reflect a normal sunny day as defined in Table I. All other LIS units are energy sources with the maximum and minimum capacity as listed in Table II. The DC Link Bus voltage reference command is set to 300 V. Switching models of DC-DC and DC-AC Converters [20] are developed in Simulink, which execute the real time power reference commands received through Serial Communication Interface (SCI) from Raspberry PI modules [21]. In this study, LIS 2 is the demand circulation node. The prioritization of RES is done by setting the capacities of LIS 2 as, $\pi^{min} = P_{MPPT}(t) - \epsilon$ and $\pi^{max} = P_{MPPT}(t)$, for all time $t$, where $\epsilon$ is a small positive number and $P_{MPPT}$ is the output power of PV obtained by implementing an MPPT algorithm at all instants. The capacities of all non-RES LIS units are listed in Table II. Agents are initialized as discussed in (5), (6), (7) with $p = 1$.

TABLE I
PV profile for experiment

| Time(Hrs)     | Power-Output(kW)  |
|---------------|-------------------|
| $0 \leq t < 3$ | ramps up to 1kW   |
| $3 \leq t < 5$ | 1kW (Constant)    |
| $5 \leq t \leq 8$ | ramps down to 0 |

Algorithm 1 is executed in a distributed manner on each R-PI unit and a power reference command is generated approximately at every second. Delays are variable and empirically delays as high as 3000 ms between R-PI units are observed

TABLE II
LIS Parameters used for validation

| Unit | $\pi_{min}(W)$ | $\pi_{max}(W)$ |
|---|---|---|
| LIS 1 | 0 | 1500 |
| LIS 2 (PV) | Profile Min(t) | Profile Max(t) |
| LIS 3 | 0 | 1000 |
| LIS 4 | 0 | 1200 |
| LIS 5 | 0 | 1500 |
| LIS 6 | 0 | 2000 |

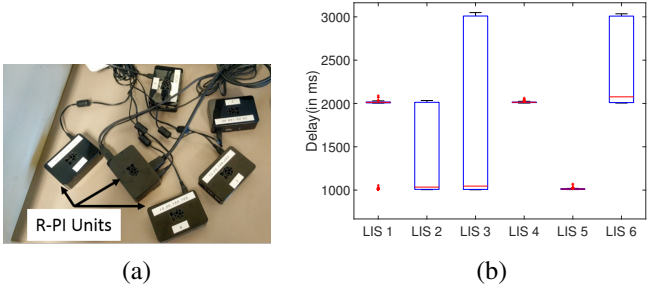

(a)      (b)

Fig. 5. (a) Raspberry PI units communicating wirelessly to their respective neighbors, (b) Pairwise latency in communication channels between R-PI units and their respective neighbors.

(see Figure 5(b)). Power reference commands are generated upon termination of Algorithm 1 at each R-PI unit based on the user-specified threshold $\rho$. Figure 6 shows successive ratio consensus cycles terminated using Algorithm 1 with each cycle taking about 30 iterations (Figure 7) for finite time termination within the given threshold. Power reference commands are then dispatched by each R-PI agent through SCI at every minute to the power electronics interface of the respective LIS unit in Simulink. The LIS units inject power into the network by following the reference power commands as shown in Figure 8. Figure 9 and Figure 10 demonstrate RES prioritization of LIS 2 and auto-adjustment of contributions from non-RES LIS units. It is evident that as more (less) energy from RES LIS 2 becomes available in the network, in order to prioritize and consume all available solar energy from LIS 2, the non-RES units (LIS 1, 3, 4, 5 and 6) auto-adjust to increase (decrease) their power injection into the network based on the ratio of their capacities. A total AC power output of the six LIS units is 7000 W ±150 W (not exactly 7000 W due to termination of the algorithm based on the threshold) at all instants as shown in Figure 11 which is equal to the ancillary services demanded by the grid. The DC Link voltages

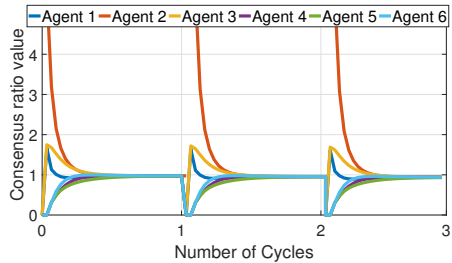

Fig. 6. Successive cycles of R-PI agents showing convergence of consensus ratio in finite time

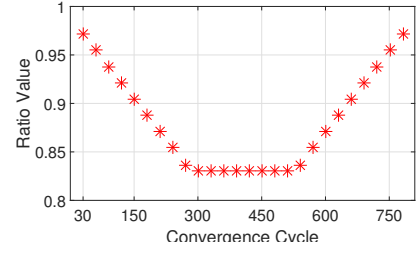

Fig. 7. Converged ratio values of Non-RES LIS units demonstrating adjustment of contributions allowing RES penetration in the network with higher priority.

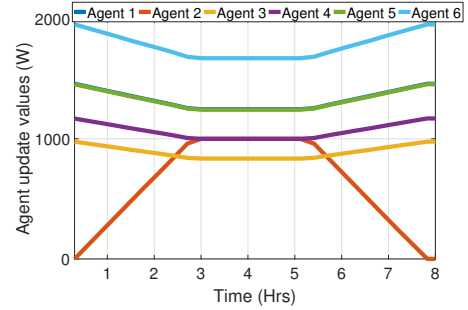

Fig. 8. Updated $\pi^*$ values for each R-PI agent after convergence of consensus iterations dispatched as Reference Power Commands to LIS unit.

of all LIS units is observed to be within the safe limits at all time.

In summary, we applied the algorithm proposed for coordination of LIS units on hardware that communicate through wireless channels that suffers from uncertainties, such as delays, to meet the grid demand. Thus, we now have a coordination algorithm which is distributed and suitable for real time application.

## 6. CONCLUSION

This paper extends the results of finite-time termination of ratio consensus algorithm to an application of distributed apportioning of load in the presence of communication delays for providing demand response services requested by the grid. The most notable result of this paper is the completely distributed nature of the resource apportioning algorithm, from selection of communication weights to the finite time termination criteria. This framework enables unique and simultaneous, yet local and independent realization of power reference commands by the distributed control units and a subsequent coordinated action. The algorithm proposed in this paper also enables prioritization of Renewable Energy Sources available in the power network and auto-adjustment of energy contributions from other non-RES units to allow for increased renewable penetration. The proposed algorithm also performs efficiently with time-varying ancillary service requests of the grid within the limits of convergence of each consensus cycle. The efficacy of the algorithm is demonstrated by hardware-in-the-loop experiments using a network of Raspberry Pi agents as distributed control units, which dispatch the power reference commands generated by the resource apportioning algorithm to the local power electronics interface.

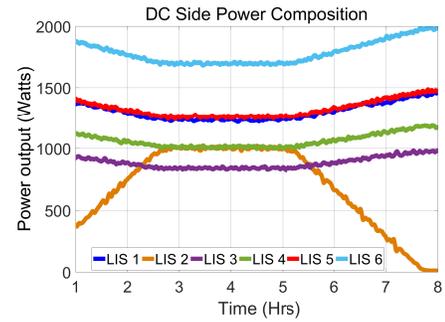

Fig. 9. DC Power Output of each LIS unit and RES Prioritization of LIS 2 by self-adjustment of contributions from non-RES LIS units.

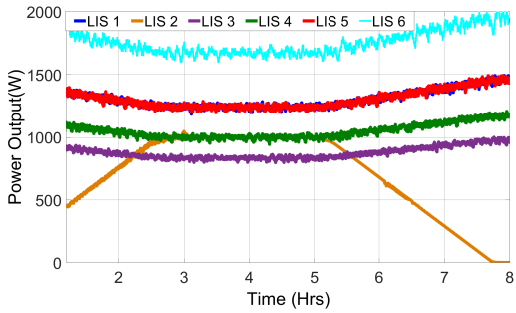

Fig. 10. AC Side Power Composition of LIS Units

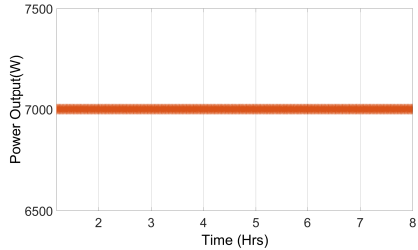

Fig. 11. Total power output from LIS units delivered in response to ancillary services requested by the grid.

ACKNOWLEDGMENT

The authors acknowledge Advanced Research Projects Agency-Energy (ARPA-E) for supporting this research through the project titled 'A Robust Distributed Framework for Flexible Power Grids' via grant no. DE-AR000071 and Xcel Energy's Renewable Development Fund.

REFERENCES

[1] R. H. Lasseter and P. Piagi, "Microgrid: A conceptual solution," in *Proc. Power Electronics Specialists Conf*, vol. Vol. 6, pp. 4285–4290, Aachen, Germany, Jun 2004.
[2] G. Joos, B. T. Ooi, D. McGillis, F. D. Galiana, and R. Marceau, "The potential of distributed generation to provide ancillary services," *Proc. IEEE Power Engineering Society Summer Meeting*, vol. Vol. 3, pp. 1762–1767, Seattle, 2000.
[3] G. Pepermans, J. Driesen, D. Haeseldonckx, R. Belmans, and W. D'haeseleer, "Distributed generation: definition, benefits and issues," *Energy Policy*, vol. 33, pp. 787–798, April 2005.
[4] E. Unamuno and J. A. Barrena, "Hybrid ac/dc microgrids part ii: Review and classification of control strategies," *Renewable and Sustainable Energy Reviews*, vol. 52, pp. 1123–1134, 2015.
[5] E. Planas, A. Gil-de Muro, J. Andreu, I. Kortabarria, and I. Martinez de Alegria, "General aspects, hierarchical controls and droop methods in microgrids: A review," *Renewable and Sustainable Energy Review*, vol. 17, pp. 147–159, 2013.
[6] A. D. Dominguez-Garcia and C. N. Hadjicostis, "Coordination and control of distributed energy resources for provision of ancillary services," in *Smart Grid Communications (SmartGridComm), 2010 First IEEE International Conference on*. IEEE, 2010, pp. 537–542.
[7] S. T. Cady, A. D. Domínguez-García, and C. N. Hadjicostis, "Finite-time approximate consensus and its application to distributed frequency regulation in islanded ac microgrids," in *System Sciences (HICSS), 2015 48th Hawaii International Conference on*. IEEE, 2015, pp. 2664–2670.
[8] V. Yadav and M. V. Salapaka, "Distributed protocol for determining when averaging consensus is reached," in *45th Annual Allerton Conf*, 2007, pp. 715–720.
[9] M. Prakash, S. Talukdar, S. Attree, S. Patel, and M. Salapaka, "Distributed finite time termination of ratio consensus for averaging in the presence of delays," *arXiv preprint arXiv:1704.08297*, 2017.
[10] J. W. Stahlhut, T. J. Browne, G. T. Heydt, and V. Vittal, "Latency viewed as a stochastic process and its impact on wide area power system control signals," *IEEE Transactions on Power Systems*, vol. 23, pp. 84–91, 2008.
[11] R. Diestel, *Graph theory {graduate texts in mathematics; 173}*. Springer-Verlag Berlin and Heidelberg GmbH & amp, 2000.
[12] R. A. Horn and C. R. Johnson, Eds., *Matrix Analysis*. New York, NY, USA: Cambridge University Press, 1986.
[13] S. Salapaka, B. Johnson, B. Lundstrom, S. Kim, S. Collyer, and M. Salapaka, "Viability and analysis of implementing only voltage-power droop for parallel inverter systems," *IEEE Conference on Decision and Control*, vol. 53, no. 3, pp. 3246–3251, 2014.
[14] M. Baranwal, S. Salapaka, and M. Salapaka, "Robust decentralized voltage control of dc-dc converters with applications to power sharing and ripple sharing," *American Control Conference*, pp. 7444–7449, 2016.
[15] C. N. Hadjicostis and T. Charalambous, "Average consensus in the presence of delays in directed graph topologies," *IEEE Transactions on Automatic Control*, vol. 59, no. 3, pp. 763–768, 2014.
[16] M. Prakash, S. Talukdar, S. Attree, V. Yadav, and M. Salapaka, "Distributed finite time termination of consensus in presence of delays," *IEEE Transactions of Automatic Control (in review)*, (arXiv preprint arXiv:1701.0002), 2016.
[17] "Raspberry pi 3 model b , https://www.raspberrypi.org/products/raspberry-pi-3-model-b/."
[18] M. J. Lee, J. Zheng, Y.-B. Ko, and D. M. Shrestha, "Emerging standards for wireless mesh technology," *IEEE Wireless Communications*, vol. 13, no. 2, pp. 56–63, 2006.
[19] R. Faranda and S. Leva, "Energy comparison of mppt techniques for pv systems," *WSEAS transactions on power systems*, vol. 3, no. 6, pp. 446–455, 2008.
[20] N. Mohan and T. M. Undeland, *Power electronics: converters, applications, and design*. John Wiley & Sons, 2007.
[21] A. Kurniawan, *Getting Started with Matlab Simulink and Raspberry Pi*. PE Press, 2013.